\documentclass[10pt]{article}

\usepackage{amsfonts}
\usepackage{amssymb}
\usepackage{amsmath}
\usepackage{amsthm}
\usepackage{amsfonts}
\usepackage{multicol}
\usepackage{amscd}
\usepackage{textcomp}
\usepackage{multirow}
\usepackage{color}
\usepackage[english, activeacute]{babel}
\usepackage{epic,eepic,graphicx}

\def\be{\begin{equation}}
\def\ee{\end{equation}}
\def\bea{\begin{eqnarray}}
\def\eea{\end{eqnarray}}

\def\tfrac#1#2{ {\scriptstyle { \frac {#1}{#2}}}}         
\def\pois#1#2{\left\{ {#1},{#2} \right\}}

\def\1{\'{\i}}

\def\>#1{{\mathbf#1}}

\begin{document}

\thispagestyle{empty}

\hfill \today

\ 
\vspace{0.5cm}

\begin{center}

{\LARGE{\sc{Integrable deformations of R\"ossler and Lorenz

systems from Poisson--Lie groups
}} }

\end{center}

\begin{center} {\sc \'Angel Ballesteros$^a$, Alfonso Blasco$^a$ and Fabio Musso$^{a,b}$}
\end{center}

\begin{center} {$^a$\it{Departamento de F\1sica,  Universidad de Burgos, 
09001 Burgos, Spain}}

{$^b$\it{I.C. ``Leonardo da Vinci",
Via della Grande Muraglia 37,
 I-0014 Rome, Italy}}

e-mail: angelb@ubu.es, ablasco@ubu.es, fmusso@ubu.es
\end{center}

  \medskip

\begin{abstract} 
A method to construct integrable deformations of Hamiltonian systems of ODEs endowed with Lie--Poisson symmetries is proposed by considering Poisson--Lie groups as deformations of Lie--Poisson (co)algebras. Moreover, the underlying Lie--Poisson  symmetry  of the initial system of ODEs is used to construct integrable coupled systems,  whose integrable deformations can be obtained through the construction of the appropriate Poisson--Lie groups that deform the initial symmetry. The approach is applied in order to construct integrable deformations of both uncoupled and coupled  versions of certain integrable types of R\"ossler and Lorenz systems. It is worth stressing that such deformations are of non--polynomial type since they are obtained through an exponentiation process that gives rise to the Poisson--Lie group from its infinitesimal Lie bialgebra structure. The full deformation procedure is essentially algorithmic and can be computerized to a large extent.
\end{abstract}

\bigskip\bigskip\bigskip\bigskip


\bigskip


\vfill
\newpage


\section{Introduction}

The integrability problem for systems of first order ODEs is indeed a relevant issue in the theory of dynamical systems (see, for instance,~\cite{Goriely, LR, Olver, Fomenko, Zakharov} and references therein), and the same question for coupled systems of ODEs could be also meaningful from the viewpoint of applications (for instance, regarding synchronization problems~\cite{synchro}). In this context, the aim of this paper is two fold: on one hand, to present a novel approach to the explicit construction of Liouville--integrable coupled systems of ODEs endowed with Lie--Poisson symmetries and, on the other hand, to show how integrable deformations of the previous systems can be systematically obtained.

The construction here presented establishes a (to the best of our knowledge) novel connection between Poisson--Lie groups and integrable deformations of dynamical systems. Such a link is based on the well-known result by Drinfel'd that establishes the one-to-one correspondence between Poisson--Lie groups and Lie bialgebras~\cite{DriPL}, as well as on the general construction of integrable Hamiltonian systems from Poisson coalgebras that was introduced in~\cite{BR, BRcluster, jpcs}, thus giving rise to the so-called ``coalgebra symmetry approach" to finite dimensional integrable systems (see~\cite{HHannals, LVpla} for applications of this method to H\'enon--Heiles and Lotka--Volterra systems, respectively, and~\cite{LieH} for the coalgebra--based construction of nonlinear superposition rules arising in nonautonomous Lie--Hamilton systems~\cite{LieHjde}). 

In particular, we will apply the formalism here introduced in order to obtain new integrable deformations of certain Lorenz~\cite{Lorenz} and R\"ossler~\cite{Rossler1} systems, as well as of some coupled versions of them. Nevertheless, we stress that the method presented in the paper is completely general and could be applied to any other interesting Hamiltonian systems of ODEs endowed with a Lie--Poisson symmetry.

The basics of the Poisson coalgebra approach to integrability will be summarized in the next Section. In short, we will consider a finite--dimensional integrable Hamiltonian system of ODEs that is defined through a Lie--Poisson algebra $({\cal F}(g^*),\{,\})$, where ${\cal F}(g^*)$ is the algebra of smooth functions on $g^*$ and the Lie--Poisson bracket is $\{,\}$. A set of  functions in involution will be 
given by a single Hamiltonian function ${\cal H}\in {\cal F}(g^*)$ and a number of Casimir functions ${\cal C}_k\in {\cal F}(g^*)$, $k=1,\dots,r$. Then, since Lie--Poisson algebras are endowed with a (primitive or non--deformed) coalgebra structure, 
the coalgebra symmetry approach will straightforwardly provide, by following~\cite{BR}, a coupled integrable generalization of the system.  This approach will be illustrated in section 3 through the explicit construction of an integrable coupled R\"ossler system. 

A detailed technical presentation of the construction of Poisson--Lie groups as deformations of Lie bialgebras will be given in Section 4. If we  denote by ${g}$ the abstract Lie algebra associated with the Lie--Poisson coalgebra $({\cal F}(g^*),\{,\})$, we will assume that we are able to find a cocommutator map $\delta$ compatible with ${g}$, so that
$({g},\delta)$ defines a Lie bialgebra structure~\cite{Dri,CP}. Then, it is well--known that the dual of the cocommutator $\delta$ will define a second Lie algebra structure (that we will denote by $d$), and the dual of the structure tensor for $g$ will provide a compatible cocommutator map $\gamma$ for $d$, in  such a way that $(d,{\gamma})$ is just the dual Lie bialgebra to $({g},\delta)$.  Once a given $\delta^\ast\simeq d$ is identified, the (connected component containing the identity) of the Lie group $D$ with Lie algebra $d$ can be constructed by exponentiation, and the unique Poisson--Lie structure having $g^\ast$ as its linearization at the identity can be   computed, thus providing the deformed Poisson structure $\{,\}_\eta$ we were looking for. Simultaneously, the group law for $D$ will give us the explicit deformed coproduct map $\Delta_\eta$ on the algebra of smooth functions ${\cal{F}}(D)$. Moreover, the deformation nature of this construction arises in a more transparent way if we consider the one-parametric cocommutator $\gamma_\eta=\eta\,\gamma$, since the parameter $\eta$ will give us information concerning the orders in the deformation process that arises from the `exponentiation'\ of the tangent Lie bialgebra structure. It is worth to be stressed that the procedure that we have just sketched is algorithmic and can be made fully explicit by making use of symbolic computation packages.

Therefore, as a Poisson coalgebra, $({\cal{F}}(D),\{,\}_\eta,\Delta_\eta)$ can be interpreted as a smooth $\eta$--defor\-mation 
of $({\cal{F}}(g^*),\{,\},\Delta)$ on which the initial dynamical system with Hamiltonian function ${\cal H}$ is defined.
As a consequence, if we consider the same Hamiltonian function ${\cal H}$ on $({\cal{F}}(D),\{,\}_\eta,\Delta_\eta)$, then the associated dynamical system is just an integrable 
$\eta$--deformation of the initial system of ODEs. Moreover, the coalgebra symmetry approach~\cite{BR} guarantees that the set of involutive functions for the deformed system will be given by the deformed coproducts of the Hamiltonian ${\cal H}$ and of  the deformed Casimir functions ${\cal C}^{\eta}_k$ for the Poisson structure $\{,\}_\eta$, with $k=1, \dots,r$. 

Sections 5 and 6 are devoted to use this Poisson-deformation approach in order to obtain integrable deformations of two coupled R\"ossler and Lorenz systems, respectively. It is worth stressing that, as a result of the exponentiation process that underlies the full construction, such integrable deformations of dynamical systems are provided by non--polynomial vector fields, which is a non--standard finding from the viewpoint of integrability. Also, the underlying Poisson--Lie groups are responsible of the fact that the coupling between the systems is quite involved, since such coupling is generated by their non--abelian group law. A final Section including some remarks and open problems closes the paper.


\section{Lie--Poisson coalgebras}

Let $g$ be a finite dimensional Lie algebra. As it is well known, the associated Lie--Poisson algebra ${\cal F}(g^*)$ is the algebra of $C^\infty$ functions on $g^*$, equipped with the Poisson bracket
\begin{equation}
\{f,g\}(x)=\langle [df,dg],x \rangle, \qquad x \in g^*, \quad f,g \in {\cal F}(g^*), \label{LPintrinsic}
\end{equation}
where $\langle, \rangle$ is the non-degenerate pairing between $g$ and $g^*$.
Let us assume that $g$ is generated by the set $\{X_i \}_{i=1}^{{\rm dim}(g)}$, with commutation relations
\begin{equation}
[X_i,X_j]=c_{ij}^k\, X_k, \label{structurecoeffs}
\end{equation}
and let us denote with $\{x_i\}_{i=1}^{{\rm dim}(g)}$ the corresponding dual basis in $g^\ast$
\begin{equation}
\langle x_i, X_j  \rangle=\delta_{ij},
\qquad i,j=1,\dots,\mbox{dim}(g).
\end{equation}
Then the Lie--Poisson bracket (\ref{LPintrinsic}) for the coordinate functions on $g^*$ is given by the fundamental Poisson brackets
\begin{equation}
\{x_i,x_j\}=c_{ij}^k \,x_k  \label{fundPoiss}.
\end{equation}

Now, by defining the following coproduct map $\Delta: {\cal{F}}(g^*) \to {\cal{F}}(g^*) \otimes {\cal{F}}(g^*)$ for the coordinate functions 
\begin{equation}
\Delta(x_i)=x_i \otimes 1 + 1 \otimes x_i , \label{primitive}
\end{equation}
we can say that $({\cal{F}}(g^*),\{,\},\Delta)$ is endowed with a Poisson coalgebra structure~\cite{BR}, since the coassociativity condition
\begin{equation}
(\Delta\otimes id)\circ \Delta = (id\otimes \Delta)\circ \Delta,
\end{equation}
is fulfilled, and the coproduct map $\Delta$ can be straightforwardly shown to define a Poisson map between ${\cal{F}}(g^*)$ and ${\cal{F}}(g^*)\otimes {\cal{F}}(g^*)$, namely
\begin{equation}
\Delta(\{f,g\})=\{\Delta(f),\Delta(g)\}, \qquad \forall f,g \in {\cal{F}}(g^*), \label{homorphism}
\end{equation}
with respect to the Poisson structure on ${\cal{F}}(g^*) \otimes {\cal{F}}(g^*)$ induced from~\eqref{fundPoiss}.

The coalgebra approach to integrability is based on the fact that if we start from a Liouville integrable system
defined on ${\cal{F}}(g^*)$, we can use the coproduct map (\ref{primitive}) in order to extend it to an integrable system defined on 
$[{\cal{F}}(g^*)]^N:= {\cal{F}}(g^*) \otimes  \dots^{N}\otimes {\cal{F}}(g^*)$, with $N$  being an arbitrary number of copies of ${\cal F}(g^*)$. The keystone for this result is given by the $k$-th coproduct maps, which are defined recursively as 
\begin{equation}
\Delta^{(k)} \equiv (\Delta \otimes id) \circ \Delta^{(k-1)}, \quad k=3,\dots,N, \qquad \Delta^{(2)} \equiv \Delta,
\label{rec}
\end{equation}
and, by construction, satisfy the Poisson homorphism property (\ref{homorphism})
\begin{equation}
\Delta^{(k)}(\{f,g\})=\{\Delta^{(k)}(f),\Delta^{(k)}(g)\}, \qquad \forall f,g \in {\cal{F}}(g^*). \label{homorphism}
\end{equation}

The simplest possible case in this framework occurs when the integrable system on ${\cal{F}}(g^*)$ is defined by a single Hamiltonian function ${\cal H}$ and 
an arbitrary number of Casimir functions ${\cal C}_k$, $k=1,\dots,r$. In this case all the involutive functions defining an integrable system on $[{\cal{F}}(g^*)]^N$
are obtained as images of the Hamiltonian and Casimir functions under the $\Delta^{(k)}$ maps $(k=2,\dots,N)$. 
In this paper we will consider only this case, but we recall that this approach can be generalized for Poisson algebras in which integrability involves either more than a single Hamiltonian or even non-coassociative Poisson maps (see~\cite{Marmo, comodule, loopsAIP}).


\section{Integrable R\"ossler systems}

In order to illustrate this integrability approach, let us consider the specific Lie--Poisson algebra that underlies the integrability of a dynamical system of R\"ossler type.
We recall that the so--called  R\"ossler system 
\begin{equation}
\begin{array}{l}
\dot{x}=-y-z,\\
\dot{y}=x+a y,\\
\dot{z}=x z+b-c  z,
\end{array}\label{S1}
\end{equation}
was introduced in \cite{Rossler1} as a prototype of 3D system of ODEs exhibiting the phenomenon of continuous chaos. 
When $a=b=c=0$, the system (\ref{S1}) is known to be integrable in the Liouville sense \cite{LlibreZhang,Llibre-Valls,LimaLLibre,TudGir}, and gives rise to 
\begin{equation}
\begin{array}{l}
\dot{x}=-y-z,\\
\dot{y}=x,\\
\dot{z}=x\,z.
\end{array}\label{RS}
\end{equation}
In particular, this system is known to be Hamiltonian~\cite{TudGir} with respect to the Poisson bracket
\begin{equation}
\begin{array}{llll}
\{x,y\}=-1, & \{x,z\}=-z, & \{y,z\}=0, \label{Tud}
\end{array}
\end{equation}
provided that we consider the Hamiltonian function
\begin{equation}
\mathcal{H}=\dfrac{1}{2}(x^{2}+y^{2})+z. \label{RH}
\end{equation}
The bracket~\eqref{Tud} can be thought of as a particular case of the Lie--Poisson bracket
\begin{equation}
\begin{array}{llll}
\{x,y\}_0=-w, & \{x,z\}_0=-z, & \{y,z\}_0=0, &\{w,\cdot \}_0=0, \label{LA}
\end{array}
\end{equation}
where we have introduced a fourth ``central" Poisson algebra generator $w$. It is immediate to show that this is just the Lie--Poisson algebra corresponding to the 4D real solvable Lie algebra $g\equiv A_{4,3}$ (here we follow the classification of four--dimensional real Lie algebras given in~\cite{invariants}). The Casimir functions for (\ref{LA}) are given by
\begin{equation}
\begin{array}{ll}
\mathcal{W}=w, \qquad&\mathcal{C}=z\, e^{-\frac{y}{w}}.
\label{cw}
\end{array}
\end{equation}
Indeed, if the Hamiltonian of the system is again given by (\ref{RH}), the R\"ossler system (\ref{RS}) is recovered when $\mathcal{W}=w=1$. 


\subsection{Coupled R\"ossler systems from Lie--Poisson coalgebras}

The Lie--Poisson coalgebra $({\cal{F}}(g^*),\{,\}_0,\Delta)$ is defined through the primitive coproduct map
\begin{eqnarray}
\Delta(x)&=&x \otimes 1+1\otimes x = x_1+x_2 ,\nonumber \\
\Delta(y)&=&y \otimes 1+1\otimes y = y_1+y_2 ,\label{coproducts}\\
\Delta(z)&=&z \otimes 1+1\otimes z = z_1+z_2 ,\nonumber \\
\Delta(w)&=&w \otimes 1+1\otimes w = w_1+w_2 ,\nonumber 
\end{eqnarray} 
where in the second equality we have identified the coordinate functions on each copy of ${\cal{F}}(g^*)$ through a different subscript.
Now, by using the standard Poisson coalgebra structure~\eqref{LA}--\eqref{coproducts}, we are able to define integrable  coupled generalizations of
the R\"ossler system (examples of nonintegrable couplings can be found in~\cite{Yanchuk, Pravitha}). Explicitly, if we consider the coproduct of the Hamiltonian~\eqref{RH}, we get
\begin{equation}
\Delta(\mathcal{H})=\dfrac{1}{2}\left([\Delta(x)]^{2}+[\Delta(y)]^{2}\right)+\Delta(z)=
\dfrac{1}{2}(x_1^{2}+y_1^{2})+z_1+\dfrac{1}{2}(x_2^{2}+y_2^{2})+z_2+x_1x_2+y_1y_2,
\end{equation}
and the corresponding 6D equations of motion lead to a coupled R\"ossler system
\begin{equation}
\begin{array}{l}
\dot{x}_{1}=-w_{1}(y_{1}+y_{2})-z_{1},\\
\dot{y}_{1}=w_{1}(x_{1}+x_{2}),\\
\dot{z}_{1}=z_{1}(x_{1}+x_{2}),\\[4pt]
\dot{x}_{2}=-w_{2}(y_{1}+y_{2})-z_{2},\\
\dot{y}_{2}=w_{2}(x_{1}+x_{2}),\\
\dot{z}_{2}=z_{2}(x_{1}+x_{2}),
\end{array}
\label{coro}
\end{equation}
together with $\dot{w}_{1}=\dot{w}_{2}=0$.
Notice that the coupling between both systems comes from the non--linear terms of the Hamiltonian~\eqref{RH}, and the restriction of the dynamics to the submanifold $x_2=y_2=z_2=w_2=0, \ w_1=1$ (or, alternatively, to $x_1=y_1=z_1=w_1=0, \ w_2=1$) leads us to the uncoupled R\"ossler system (\ref{RS}). 

The 6D Hamiltonian system~\eqref{coro} is integrable since, due to the underlying coalgebra symmetry, we have three additional (besides the Hamiltonian) independent integrals of the motion that are in involution. Two of them are given by the (nontrivial) Casimir functions on each of the copies of the Lie--Poisson algebra, namely
\be
\mathcal{C}_1=z_1\, e^{-\frac{y_1}{w_1}},
\qquad
\mathcal{C}_2=z_2\, e^{-\frac{y_2}{w_2}},
\ee
and the the third one is just the coproduct of the Casimir $\mathcal{C}$, which reads
\begin{equation}
\Delta(\mathcal{C})=\Delta(z)\, e^{-\frac{\Delta(y)}{\Delta(w)}}=(z_1+ z_2)\, e^{-\frac{(y_1+ y_2)}{(w_1+ w_2)}}.\end{equation}

The $N$--dimensional generalization of this result is straightforward by considering the $N$-th coproduct map $\Delta^{(N)}$ of the Hamiltonian and of the Casimir functions, that is obtained as the pull-back through the $N$-th coproduct of the coordinate functions~\cite{BR}
\bea
&&\!\!\!\!\!\!\!\!\Delta^{(N)}(x)=x\otimes 1\otimes 1\otimes
\dots^{N-1)}\otimes 1 \cr
&&\qquad\qquad\quad + 1\otimes x\otimes 1\otimes\dots^{N-2)}\otimes
 1 +
\dots \cr
&&\qquad\qquad\qquad\quad + 1\otimes
1\otimes\dots^{N-1)}\otimes 1\otimes x,
\label{fo}
\eea
(and the same for $y$ and $z$). Explicitly, 
\begin{equation}
H^{(N)}:=\Delta^{(N)}(\mathcal{H})=\dfrac{1}{2}\left([\Delta^{(N)}(x)]^{2}+[\Delta^{(N)}(y)]^{2}\right)+\Delta^{(N)}(z)=
\dfrac{1}{2}\left(\sum_{i=1}^N x_i\right)^{2} + \dfrac{1}{2}\left(\sum_{i=1}^N y_i \right)^{2}+\sum_{i=1}^N z_i,
\label{NR}
\end{equation}
and the $N$-coupled R\"ossler system reads
\begin{equation}
\begin{array}{l}
\dot{x}_{i}=-w_{i}\left(\sum_{j=1}^N y_j\right)-z_{i},\\
\dot{y}_{i}=w_{i}\left(\sum_{j=1}^N x_j\right),\qquad\qquad\qquad i=1,\dots,N\\
\dot{z}_{i}=z_{i}\left(\sum_{j=1}^N x_j\right).
\end{array}
\label{coroN}
\end{equation}
This system of $3\,N$ coupled ODEs has the following ($2N -1 $) integrals of the motion:
\be
\mathcal{C}_i=z_i\, e^{-\frac{y_i}{w_i}}, \qquad i=1,\dots,N
\qquad
\ee
\begin{equation}
C^{(k)}:=\Delta^{(k)}(\mathcal{C})=\Delta^{(k)}(z)\, e^{-\frac{\Delta^{(k)}(y)}{\Delta^{(k)}(w)}}=
\left( \sum_{i=1}^k z_i \right)\,e^{-\frac{\left( \sum_{i=1}^k y_i \right)}{\left( \sum_{i=1}^k w_i \right)}},
\qquad
k=2,\dots, N.
\end{equation}
By construction, all these functions are in involution and Poisson-commute with the Hamiltonian $H^{(N)}$.


\section{Lie bialgebras and Poisson--Lie groups}

The aim of this paper is to construct integrable deformations of coalgebra symmetric systems of the type~\eqref{coroN} for any number $N=1,2,\dots$ of coupled constituents. As we will show in the sequel, this can be achieved by considering the Lie--Poisson algebra $g^\ast$ as the linearization of a Poisson--Lie group structure on a certain {\em different} Lie group $D$. In his way, the construction of the corresponding full Poisson--Lie group structure $({\cal{F}}(D),\{,\}_\eta,\Delta_\eta)$ will provide a Poisson coalgebra $\eta$--deformation of $({\cal{F}}(g^*),\{,\}_0,\Delta)$, and the application of the coalgebra approach to a Hamiltonian defined on the Poisson--Lie group $D$ will automatically provide an integrable $\eta$--deformation of the initial system defined on ${\cal{F}}(g^*)$. 


\subsection{Lie bialgebras}

Given a Lie algebra $g$ with basis $\{X_i \}_{i=1}^{{\rm dim}(g)}$ and Lie bracket
\begin{equation}
[X_i,X_j]=c_{ij}^k\, X_k, \label{structurecoeffs}
\end{equation}
a Lie bialgebra structure $(g,\delta)$ is given by a skew-symmetric cocommutator map $\delta: g\rightarrow g \otimes g$
\begin{equation}
\delta(X_k)=f^{ij}_k\, X_i \wedge X_j, \label{cocommutator}
\end{equation}
such that\begin{itemize}
\item i) $\delta$ is a 1-cocycle, {\em  i.e.},
\be
\delta([X,Y])=[\delta(X),\,  Y\otimes 1+ 1\otimes Y] + 
[ X\otimes 1+1\otimes X,\, \delta(Y)] ,\qquad \forall \,X,Y\in
{g}.
\label{1cocycle}
\ee
\item ii) The dual map $\delta^\ast:{g}\otimes {g} \to
{g}$ is a Lie bracket.
\end{itemize}
Therefore, the dual $\delta^\ast$ of the cocommutator map defines a second (in general, different) Lie algebra structure, that we will call $d$, and whose Lie brackets are
\begin{equation}
[\hat x^i,\hat x^j]= f^{ij}_k \, \hat x^k. \label{geta} 
\end{equation}   
Here the duality relation is established through the canonical pairing $\langle  \hat x^j,X_k \rangle=\delta_k^j$. Note that, given the initial Lie algebra $g$, the dual Lie algebra $d$  has to fulfill  the 1-cocycle condition, which implies the following compatibility equations between the two Lie algebras $g$ and $d$:
\be
f^{ab}_k c^k_{ij} = f^{ak}_i c^b_{kj}+f^{kb}_i c^a_{kj}
+f^{ak}_j c^b_{ik} +f^{kb}_j c^a_{ik}. 
\label{compatfc}
\ee

By taking the components of the structure tensor $c$ as initial data, the solutions to the (now linear) equations~\eqref{compatfc} together with the (quadratic) equations coming from the Jacobi identity for $f$ provide all possible Lie bialgebra structures for $g$. Finally, if the solutions for the tensor $f^{ij}_k$ are classified into equivalence classes under the action of all possible automorphisms for $g$, the classification of all possible Lie bialgebra structures of $g$ is obtained.

Two remarks are in order:
\begin{itemize}

\item It is immediate to prove that if $(g,\delta)$ is a Lie bialgebra with structure tensors $(c,f)$, then $(d,\gamma)$ with $\gamma: d\rightarrow d\otimes d$ being the cocommutator map
\begin{equation}
\gamma(\hat x^k)=c_{ij}^k \,\hat x^i \wedge \hat x^j, \label{dcocommutator}
\end{equation}
defines a Lie bialgebra structure on $d$ with structure tensors $(f,c)$, that is called the dual Lie bialgebra to $(g,\delta)$. In other words, $d$ and $\delta$ are defined through the same structure tensor $f$, and we can say that $\delta^\ast\equiv d$. In the same manner,
$g$ and $\gamma$ are defined by the same structure tensor $c$ and we can say that $\gamma^\ast\equiv g$.

\item For our purposes it will be convenient to multiply all the components $f^{ij}_k$ of the cocommutator map by a real parameter $\eta$, that will play the role of a deformation parameter. The resulting cocommutator will be denoted by $\delta_\eta$,
and the dual  Lie algebra structure $d_\eta$  will be 
\begin{equation}
[\hat x^i,\hat x^j]=\eta f^{ij}_k \,\hat x^k. \label{geta} 
\end{equation}   
Obviously, $d$ is just $d_{\eta=1}$ and these two Lie algebras are isomorphic under the change of basis $\hat x^i\rightarrow \eta\,\hat x^i$. Nevertheless, as we will see in the sequel, the explicit appearance of $\eta$ will allow us to control the full deformation/exponentiation procedure.

\end{itemize}


\subsection{Poisson--Lie groups and Drinfel'd theorem}

A Poisson--Lie (PL) group is a Lie group $D$ endowed with a Poisson structure on the algebra $C^\infty(D)$ of functions on the group manifold, such that the Lie group multiplication is a Poisson map. Therefore, by defining the coproduct $\Delta(f)$ as
the pull back of a function $f\in C^\infty(D)$ by the group multiplication, we obtain a natural structure of Poisson coalgebra on the algebra of smooth functions on the Poisson--Lie group ${\cal{F}}(D)$.   
In algebraic terms, if $D$ is a Lie group, the group multiplication on $D$ induces a coproduct map
\be
\Delta: C^\infty(D)\rightarrow C^\infty(D)\otimes  C^\infty(D),
\ee
which is coassociative (since the group law is associative).
Therefore, PL groups are instances of Poisson coalgebras, since the coproduct map is -by definition- a Poisson algebra homomorphism:
\be
\pois{\Delta (a)}{\Delta (b)}=\Delta(\pois{a}{b}) \qquad
\forall\,a,b\,\in C^\infty(D),
\ee
where
\be
\{ a \otimes b, c \otimes d\}=\{a,c \} \otimes bd +ac \otimes \{ b, d \},
\label{pois2}
\ee
is the natural Poisson structure on $C^\infty(D)\otimes  C^\infty(D)$.

Now, the relevance of Lie bialgebras arises due to their one-to-one correspondence with Poisson--Lie groups, a result that can be explicitly stated as follows.

\noindent {\bf Theorem~\cite{DriPL}}. {\em 
Let $D$ be a Lie group with Lie algebra $d$:

\noindent a) If $(D,\{,\})$ is a Poisson--Lie group, then $d$ has a natural Lie bialgebra
structure $(d,\gamma)$, called the tangent Lie bialgebra of $(D,\{,\})$.

\noindent b) Conversely, if $D$ is connected and simply connected, every Lie bialgebra 
structure $(d,\gamma)$ is the tangent Lie bialgebra of a unique Poisson structure on $D$ which makes $D$ into a Poisson--Lie group.}

Explicitly,  if $(D,\{,\})$ is a PL group, the canonical Lie algebra structure on $d^\ast$ is given by
\be
[\xi^1,\xi^2]_{d^\ast}:=(d\{f_1,f_2\})_e=\gamma^\ast ( \xi^1 \otimes \xi^2), \qquad \xi^1,\xi^2 \in d^\ast
\ee
where $f_1,f_2 \in C^\infty(D)$ are such that
$(df_i)_e=\xi^i$. In other words, the linearization of the PL bracket in terms of the local canonical coordinates $\xi^i$ on the $D$ group manifold is just $\gamma^\ast$, the dual of the cocommutator map that defines the associated tangent Lie bialgebra structure.


\subsection{Poisson--Lie groups as deformations of Lie--Poisson coalgebras}

A novel construction of coalgebra deformations of Lie--Poisson algebras can be envisaged if Drinfel'd theorem is revisited by taking the tangent Lie bialgebra $(d,\gamma)$ as the initial data for the construction of the associated PL group. Let us assume that we want to construct a Poisson coalgebra deformation of a given Lie--Poisson algebra $g^\ast$. Then, we can proceed as follows:
\begin{enumerate}
\item Take the Lie algebra $g$, find a non--trivial Lie bialgebra structure $(g,\delta)$ and construct its  dual Lie bialgebra $(d,\gamma)$.
Now, recall that Drinfel'd theorem ensures that $(d,\gamma)$ will be the tangent Lie bialgebra structure of a certain Poisson--Lie group $(D,\{,\})$ such that $d=\mbox{Lie}(D)$. 

\item Introduce the isomorphic Lie algebra $d_\eta$~\eqref{geta} in order to have an explicit deformation parameter in the tangent Lie bialgebra.

\item Construct the PL group $(D,\{,\})$, that will provide the deformation of the Poisson coalgebra $({\cal{F}}(g^*),\{,\}_0,\Delta)$.
In order to obtain it,  we have firstly to find a faithful representation $\rho$ of $d_\eta$, and 
use it to construct the matrix element of the (connected component) of the Lie group $D$ through the usual exponentiation:
\begin{equation}
D=\prod_{i=1}^{{\rm dim}(d)} \exp(x_i \,\rho(\hat x^i)). \label{matrixgroup}
\end{equation}

\item 
With an abuse of notation, we will make use the same symbol $x_i$ for the coordinate functions on $g^*$ and for the local coordinates of the $D$ group  in (\ref{matrixgroup}). 
In this way, the pullback of the group multiplication will provide a coproduct map 
\be
\Delta_\eta: {\cal{F}}(D) \to {\cal{F}}(D) \otimes {\cal{F}}(D),
\ee
where, due to the coordinate identification that we have previously made, the coproduct $\Delta_\eta$ is a $\eta$ deformation of the primitive one for $({\cal{F}}(g^*),\{,\}_0,\Delta)$, namely
\begin{equation} 
\lim_{\eta \to 0} \Delta_\eta(x_i)=x_i \otimes 1 + 1 \otimes x_i=\Delta(x_i).
 \label{Deltadeformation} 
\end{equation} 
Note that in the `non-deformed case'\  $\eta=0$ we have from~\eqref{geta} that the dual Lie algebra $d$ is $[x^i,x^j]=0$, and then the group $D$ is abelian. Therefore, its group law is just the addition of group coordinates, which is algebraically expressed through the primitive coproduct~\eqref{Deltadeformation}.

\item Therefore, by Drinfel'd theorem, there exists a unique Poisson bracket $\{,\}_\eta$ on ${\cal{F}}(D)$ and such that 
$({\cal{F}}(D),\Delta_\eta,\{,\}_\eta)$ is a Poisson--Lie group
\begin{equation}
\Delta_\eta(\{F,G\})=\{\Delta_\eta(F),\Delta_\eta(G)\}, \qquad \forall F,G \in {\cal{F}}(G^*). \label{homorphismgroup}
\end{equation}
It turns out that the Poisson bracket $\{,\}_\eta$ is also an $\eta$ deformation 
of the Lie--Poisson bracket $\{,\}$ on $g^\ast$, in the sense that
\begin{equation} 
\lim_{\eta \to 0} \{x_i,x_j\}_\eta=\{x_i,x_j\}. \label{bracketdeformation} 
\end{equation} 
Indeed, the Poisson--Lie bracket $\{,\}_\eta$ has to be explicitly found by solving simultaneously the equations~\eqref{homorphismgroup} and~\eqref{bracketdeformation}, but this problem turns out to be computationally feasible.
\end{enumerate}

Due to the properties (\ref{Deltadeformation}) and (\ref{bracketdeformation}) we can conclude that, as a Poisson coalgebra, the Poisson--Lie group
$({\cal{F}}(D),\Delta_\eta,\{,\}_\eta)$ is an $\eta$--deformation of the Lie--Poisson algebra $({\cal{F}}(g^*),\{,\},\Delta)$.
Indeed, there will be as many Poisson-coalgebra deformations of $g^\ast$ as different PL groups $D$ such that $g^\ast$ induces an admissible tangent Lie bialgebra cocommutator $\gamma$ for $d=\mbox{Lie}(D)$.



\section{An integrable deformation of the R\"ossler system}

In the sequel we will make use of the fact that the integrable R\"ossler system is an outstanding example of Lie--Poisson coalgebra-symmetric system. This will  allow the construction of an $\eta$--deformed coproduct map that will give rise to integrable coupled systems of ODEs of R\"ossler type. 

Let us start by considering the real Lie algebra A$_{4,3}$ in the basis given in~\cite{invariants} (note that this is exactly the four dimensional algebra~\eqref{LA})
\begin{equation}
\begin{array}{llll}
[X,Y]=-W, \quad & [X,Z]=-Z, \quad& [Y,Z]=0, \quad&[W,\cdot ]=0. \label{LAconm}
\end{array}
\end{equation}
It is straightforward to check that the following cocommutator map
\be
\delta_\eta(X)= \eta\, X \wedge Z ,\qquad
\delta_\eta(Y)= \eta\, Z \wedge W ,\qquad
\delta_\eta(Z)=\delta_\eta(W)= 0,
\ee
is compatible with the Lie--Poisson algebra (\ref{LA}) in the sense of~\eqref{1cocycle}, and thus defines a Lie bialgebra structure $(A_{4,3}, \delta_\eta)$. The dual Lie algebra $d_\eta$ obtained from the dual cocommutator map is
\begin{eqnarray}
&& [\hat x,\hat y]=0, \qquad [\hat x,\hat z]= \eta\, \hat x, \qquad [\hat x,\hat w]=0,  \nonumber \\
&& [\hat y,\hat z]= 0, \qquad [\hat y,\hat w]= 0, \qquad [\hat z,\hat w]= \eta\, \hat y. \label{Pda}
\end{eqnarray}
It can be checked that the dual Lie algebra~\eqref{Pda} is also isomorphic to the $A_{4,3}$ algebra. Therefore, we have a self--dual Lie bialgebra structure.

Now we have to construct the Lie group $D$ whose Lie algebra is $d_\eta$. If we define the $7\times 7$ matrix $e_i^j$ as the one with the only nonvanishing entry $1$ in the $i$--th row and $j$--th column, a faithful representation $\rho$ of the  algebra~\eqref{Pda} is given by
\begin{eqnarray}
\rho(\hat x)&=&\eta\,e_1^3, \nonumber\\
\rho(\hat y)&=&\eta\,e_5^7, \nonumber\\
\rho(\hat z)&=&-\eta\,e_1^1+\eta\,e_2^4+\eta\,e_5^6, \label{rep}\\
\rho(\hat w)&=&-\eta\,e_2^3+\eta\,e_6^7. \nonumber
\end{eqnarray}
By using $\rho$, we construct the matrix group element
\begin{eqnarray}
D&=&\exp(x \rho(\hat x))\exp(y  \rho(\hat y)) \exp(z  \rho(\hat z)) \exp(w  \rho(\hat w))= \nonumber \\
&=& \left( 
\begin{array}{ccccccc}
e^{-\eta z} & 0 & \eta x  & 0 & 0 & 0 & 0\\
0 & 1 & -\eta w & \eta z & 0 & 0 & 0 \\
0 & 0 & 1 & 0 & 0 & 0 & 0 \\
0 & 0 & 0 & 1 & 0 & 0 & 0 \\
0 & 0 & 0 & 0 & 1 & \eta z & \eta(y+\eta w z) \\
0 & 0 & 0 & 0 & 0 & 1 & \eta w \\
0 & 0 & 0 & 0 & 0 & 0 & 1
\end{array}
\right), \label{matrix}
\end{eqnarray}
where $\{x,y,z,w\}$ are the local coordinate functions on the group $D$. At this point it is worth stressing that the parameter $\eta$ appears as as multiplicative factor within all the nonvanishing matrix entries of the fundamental representation of the Lie algebra $d$~\eqref{rep}. Therefore, when we exponentiate in order to obtain the group element $D$, the powers of $\eta$ will give us the corresponding contributions coming from different powers of the Lie algebra generators. 

By solving the functional equations $\Delta_\eta(D_{ij})=\sum_{k=1}^7 D_{ik} \otimes D_{kj}$ through the algorithm presented in~\cite{dualJPA}, we get the group law, {\em i.e.}, the coproduct maps for the coordinate functions of $D$, which read
\begin{eqnarray}
\Delta_{\eta}(x)&=&  x \otimes 1 + e^{-\eta z} \otimes x=x_1 +e^{-\eta z_1} x_2 , \label{cox}\cr
\Delta_{\eta}(y)&=&  y \otimes 1 + 1 \otimes y -\eta w \otimes z=y_1+y_2 -\eta w_1 z_2 ,\label{coy}\cr
\Delta_{\eta}(z)&=&  z \otimes 1 + 1 \otimes z=z_1+z_2 ,\label{coz}\\
\Delta_{\eta}(w)&=&  w \otimes 1 + 1 \otimes w=w_1+w_2,
\nonumber
\end{eqnarray}
that are indeed $\eta$-deformations of the primitive coproduct (\ref{coproducts}) such that $\lim_{\eta\to 0} \Delta_{\eta}=\Delta$.
Again, by following the approach described in~\cite{dualJPA}, the unique Poisson--Lie bracket $\{\,,\,\}_\eta$ for which the coproduct  $\Delta_\eta$ is a Poisson map is found to be
\begin{eqnarray}
\{x,y\}_\eta=-w e^{-\eta z} , \quad \{x,z\}_\eta=\frac{e^{-\eta z}-1}{\eta}, \quad \{y,z\}_\eta=0, \quad \{w,\cdot \}_\eta=0.
\label{PLdef}
\end{eqnarray}
As expected, $\lim_{\eta\to 0} \{\cdot, \cdot \}_\eta= \{\cdot, \cdot \}_0$ and we have thus obtained a $\eta$--deformation of the Lie--Poisson algebra (\ref{LA}).
Finally, the Casimir functions for the algebra~\eqref{PLdef} are found to be
\begin{eqnarray}
\mathcal{W}_\eta&=&w ,\label{wcdef0}\\ 
\mathcal{C}_\eta&=&\frac{1-\exp(-\eta z)}{\eta} \exp\left(- \frac{y}{w} \right),
\label{wcdef}
\end{eqnarray}
and, as expected, they are again $\eta$--deformations of the Casimir functions~\eqref{cw}. 

Therefore, the Poisson--Lie group $({\cal{F}}(D),\Delta_\eta,\{,\}_\eta)$ has been fully constructed. Now, by taking the same Hamiltonian~\eqref{RH} of the R\"ossler system,  the PL bracket~\eqref{PLdef} gives rise to the system of ODEs
\begin{equation}
\begin{array}{l}
\displaystyle \dot{x}=-y w e^{-\eta z}+\left( \frac{ e^{-\eta z}-1}{\eta}\right) =- (w y+z)+\left(w y z+\frac{z^2}{2}\right) \eta -\left(\frac{1}{2} w y z^2+\frac{z^3}{6}\right) \eta ^2+o[\eta^3 ], \cr
\displaystyle \dot{y}=x w e^{-\eta z}=w x-(w x z) \eta +\left(\frac{1}{2} w x z^2\right) \eta ^2+o[\eta^3 ], \\
\displaystyle \dot{z}=x\frac{\left( 1-e^{-\eta z}  \right) }{\eta} = x z- \left(\frac{1}{2} x z^2\right) \eta +\left(\frac{1}{6} x z^3\right) \eta ^2+ o[\eta^3], \cr
\displaystyle \dot{w}=0,
\end{array} 
\label{Req}
\end{equation}
that, for $w=1$, provides an integrable deformation of the R\"ossler system~\eqref{RS}, with deformed integrals of the motion given by~\eqref{wcdef0} and \eqref{wcdef}. The preservation of the closed nature of the trajectories under deformation is clearly appreciated in Figure 1.

 
 \vspace{0.5cm}

\begin{figure}[ht]
\setlength{\unitlength}{1mm}
\begin{picture}(140,66)(0,0)
\label{figure2}
\footnotesize{
\put(11,3){\includegraphics[scale=0.25]{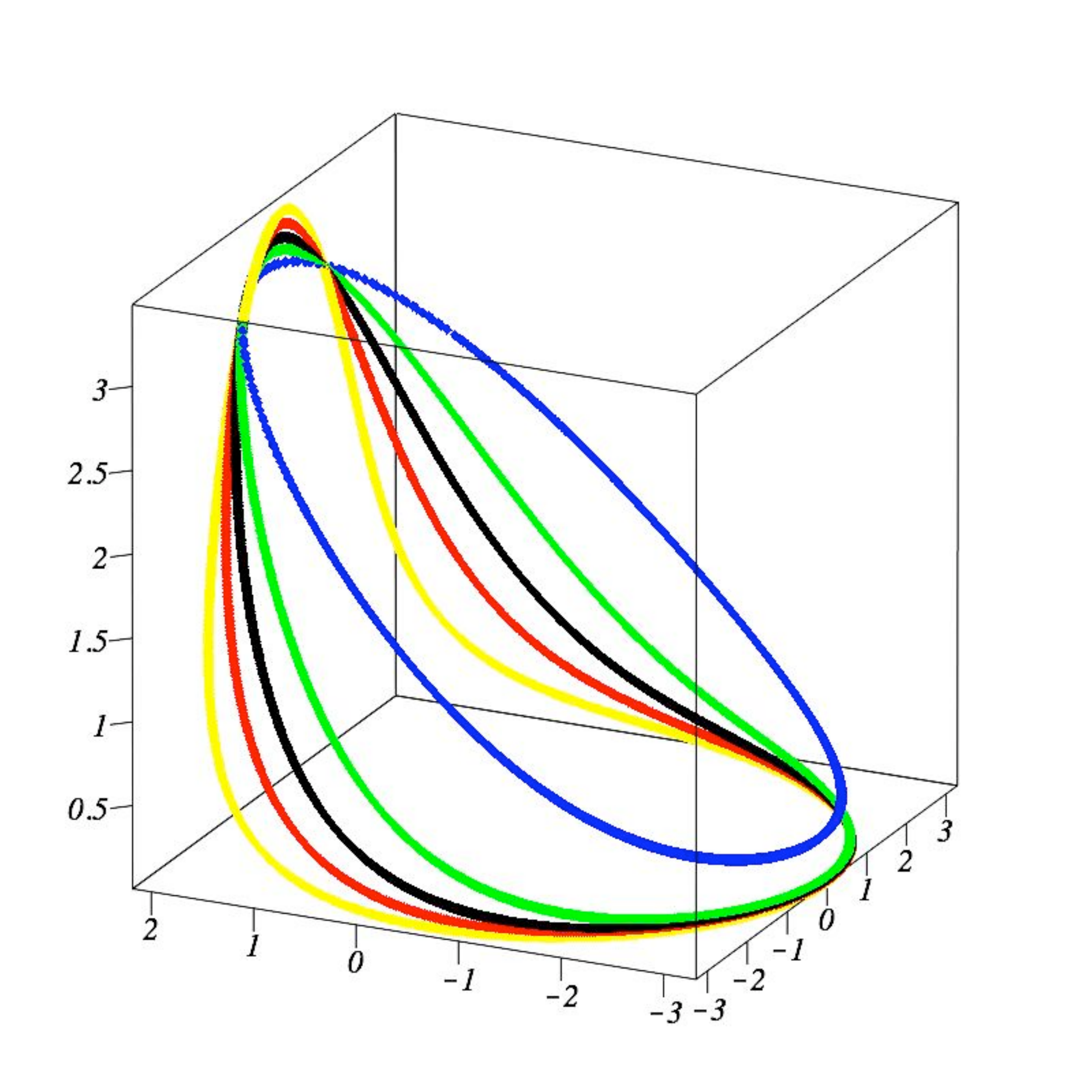}}
\put(38,6){\footnotesize $y$}
\put(68,12){\footnotesize $x$}
\put(13,36){\footnotesize $z$}
\put(11,1){(a)}
\put(85,3){\includegraphics[scale=0.25]{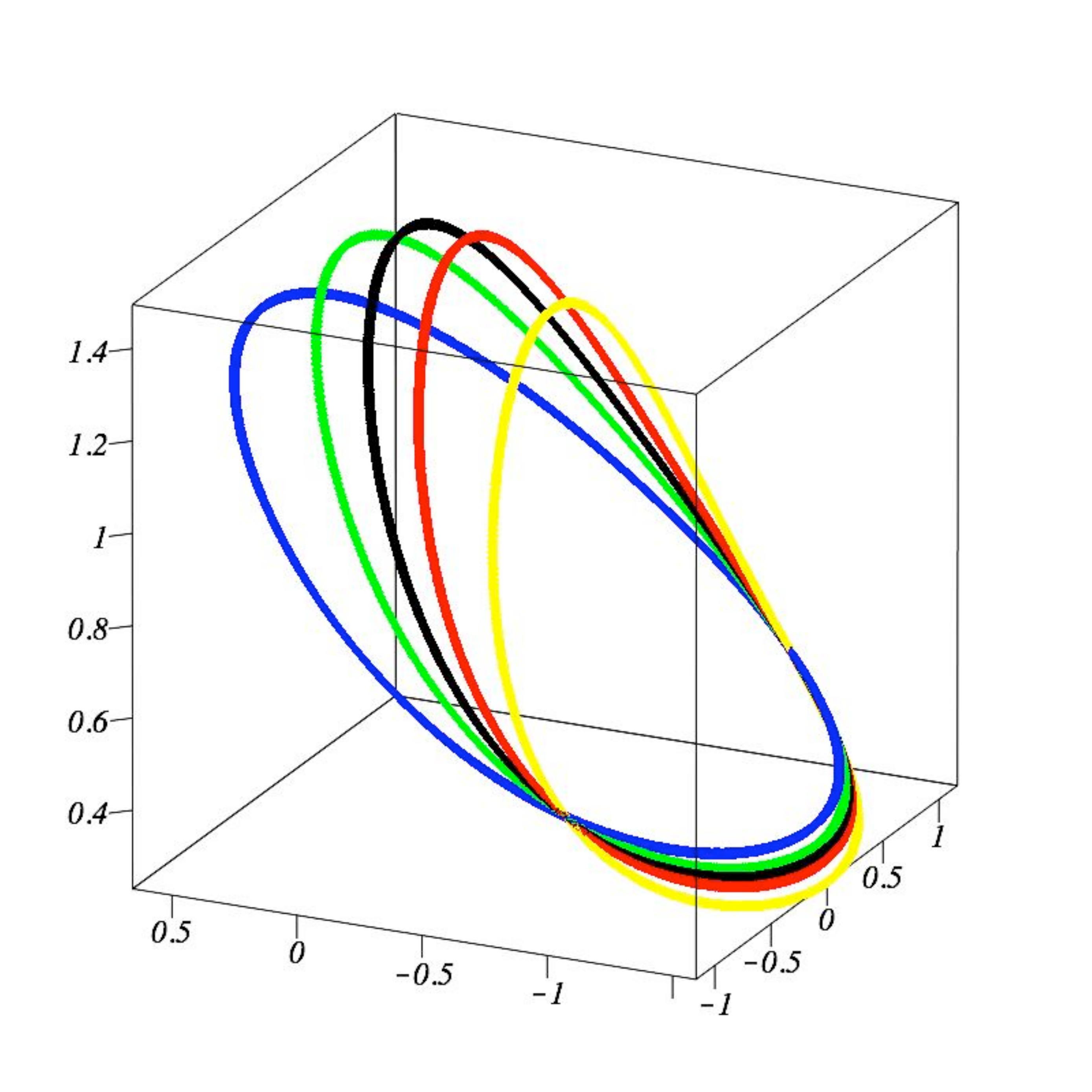}}
\put(112,6){\footnotesize $y$}
\put(142,12){\footnotesize $x$}
\put(87,36){\footnotesize $z$}
\put(85,1){(b)}
}
\end{picture}
\caption{ \footnotesize  
 (a) Closed trajectories of the integrable R\"ossler system (\ref{RS}) for the initial data $x(0)=1,y(0)=2,z(0)=3$ (black line) and of the deformed    
 integrable R\"ossler system (\ref{Req}) with the same initial data, $w=1$ and $\eta=-1.5$ (blue line), $\eta=-0.5$ (green line),
 $\eta=0.5$ (red line), $\eta=1.5$ (yellow line). (b) The same figure as (a) but with initial data $x(0)=1,y(0)=-1,z(0)=0.5$. }
\end{figure}
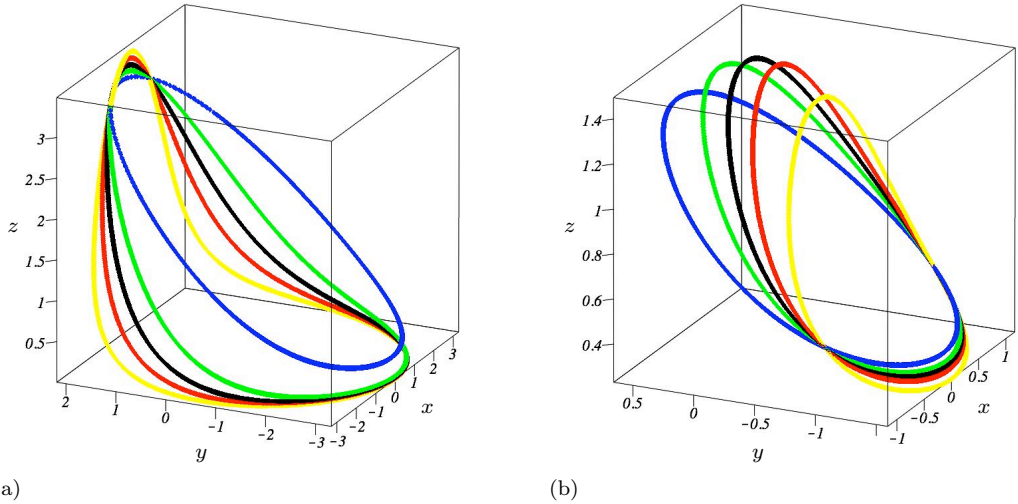


\subsection{Deformed coupled systems}

Once the Poisson--Hopf algebra $({\cal{F}}(D),\Delta_\eta,\{,\}_\eta)$ has been fully determined, the construction of an integrable deformation for the $N$--coupled system~\eqref{coroN} is straightforward. Namely, in the $N=2$ case the deformed coproduct $\Delta_\eta$ on the Hamiltonian of the R\"ossler system (\ref{RH}) will be
\be
\Delta_\eta(\mathcal{H})=\dfrac{1}{2}\left([\Delta_\eta(x)]^{2}+[\Delta_\eta(y)]^{2}\right)+\Delta_\eta(z),
\ee 
and from~\eqref{coz} we get
\begin{equation}
\begin{array}{l}
\Delta_\eta(H)=\dfrac{1}{2}\left(
(x_{1}+e^{-\eta z_{1}}x_{2})^{2}+(y_{1}+y_{2}-\eta w_{1}z_{2})^{2}
\right)+(z_{1}+z_{2})\\
=\frac12 x_1^2+x_1 x_2 e^{-\eta z_1} +\dfrac{1}{2}\left( e^{-2 \eta z_1} x_2^{2}+(y_1+y_2)(y_1+y_2-2 \eta w_1 z_2) +\eta^2 w_1^2 z_2^2\right)+ z_1+ z_2.
\end{array}
\end{equation}
This Hamiltonian gives rise to the following integrable deformation of the $N=2$ coupled R\"ossler system
\begin{eqnarray}
\dot{x_1}&=&-x_2(e^{-3 \eta z_1}-e^{-2 \eta z_1})  (x_1 e^{\eta z_1}+x_2)- \cr
&&  - w_1 (y_1+y_2-\eta w_1 z_2)e^{-\eta z_1}+\left(\frac{e^{-\eta z_1}-1}{\eta}\right),
\cr
\dot{x_2}&=&(w_1-(w_1+w_2) e^{-\eta z_2})(y_1+y_2-\eta w_1 z_2)+\left(\frac{e^{-\eta z_2}-1}{\eta}\right),\cr
\dot{y_1}&=&w_1 e^{-\eta z_1} (x_1 +x_2 e^{-\eta z_1}),\cr
\dot{y_2}&=&w_2 e^{-\eta (z_1+z_2)} (x_1 + e^{-\eta z_1} x_2),\label{2RS}\\
\dot{z_1}&=&\left(\frac{1-e^{-\eta z_1}}{\eta}\right) (x_1+x_2 e^{-\eta z_1}),\cr
\dot{z_2}&=&\left(\frac{1-e^{-\eta z_2}}{\eta}\right) e^{-\eta z_1}(x_1+x_2 e^{-\eta z_1}).
\nonumber
\end{eqnarray}
The three integrals of the motion are given by the coproduct of the Casimir $C^{(2)}_\eta$ 
\be
\Delta_\eta(C^{(2)}_\eta)=\dfrac{ [1-\exp(-\eta(z_1+z_2))]}{\eta} \exp\left(-\frac{y_1+y_2-\eta w_1 z_2}{ (w_1+w_2)} \right),
\ee
together with the `one--site'\ deformed Casimirs
\be
\mathcal{C}^{(1)}_\eta=\frac{1-\exp(-\eta z_1)}{\eta} \exp\left(- \frac{y_1}{w_1} \right),
\qquad
\mathcal{C}^{(2)}_\eta=\frac{1-\exp(-\eta z_2)}{\eta} \exp\left(- \frac{y_2}{w_2} \right).
\ee
The generalization of this construction to an arbitrary number $N$ of integrable coupled R\"ossler systems is straightformard by making use of the $N$-th order coproduct map, that can be explicitly obtained through the recurrence~\eqref{rec}.


\section{An integrable deformation of a conservative Lorenz system}

The approach described above can be applied to any other Hamiltonian system of ODEs endowed with Lie--Poisson symmetry. In particular, the celebrated Lorenz equations \cite{Lorenz} are given by the system
\begin{equation}
\begin{array}{l}
\dot{x}=\sigma(y-x), \\
\dot{y}=\rho x-x z-y, \\
\dot{z}=-\beta z+ x y.
\end{array} \label{Lorenz}
\end{equation}
Here we will consider a conservative limit of (\ref{Lorenz}) (see \cite{KAR}), which is obtained through the following rescaling
\begin{equation}
t\rightarrow \epsilon\, t,\quad
x\rightarrow \dfrac{1}{\epsilon}x, \quad
y\rightarrow \dfrac{1}{\sigma \epsilon^{2}}y, \quad
z\rightarrow \dfrac{1}{\sigma \epsilon^{2}}z, \quad
\epsilon =\dfrac{1}{\sqrt{\sigma \rho}},
\end{equation}
and by taking the limit $\rho \rightarrow \infty$.
As a consequence of this transformation, the system (\ref{Lorenz}) is mapped into
\begin{equation}
\begin{array}{l}
\dot{x}=y,\\
\dot{y}=x \,(1-z),\\
\dot{z}=x\, y,
\end{array}\label{CL}
\end{equation}
which is a Hamiltonian system with respect to the Lie--Poisson algebra \cite{KAR}
\begin{equation}
\{x,y\}=-\dfrac{1}{2} (z+1), \qquad \{x,z\}=\dfrac{y}{2}, \qquad \{y,z\}=x, \label{CLT1}
\end{equation}
and provided that we have considered the Hamiltonian function
\begin{equation}
\mathcal{H}=2(z-1)- x^{2}. \label{H}
\end{equation}

In order to proceed in a similar way as for the R\"ossler system, let us consider the 4D Lie algebra
\begin{equation}
[X,Y]=-\dfrac{1}{2} (Z+W), \qquad [X,Z]=\dfrac{1}{2} Y, \qquad [Y,Z]=X, \qquad [W,\cdot]=0. \label{lieL}
\end{equation}
This Lie algebra  is isomorphic to the $gl(2)\simeq sl(2,R)\oplus R$ Lie algebra, where $W$ is a central generator. Obviously, the Lie--Poisson algebra $g^\ast$ arising from~\eqref{lieL} is a generalization of the Poisson algebra~\eqref{CLT1} through a new central coordinate function $w$
\begin{equation}
\{x,y\}=-\dfrac{1}{2} (z+w), \qquad \{x,z\}=\dfrac{y}{2}, \qquad \{y,z\}=x, \qquad \{w,\cdot\}=0.
\label{CLT}
\end{equation}
Now, if we consider the Hamiltonian
\begin{equation}
\mathcal{H}=2(z-w)- x^{2}, \label{H1}
\end{equation}
the equations of motion obtained through~\eqref{CLT} are
\begin{equation}
\begin{array}{l}
\dot{x}=y,\\
\dot{y}=x\, (2-z-w),\\
\dot{z}=x\, y,
\end{array}\label{CL2}
\end{equation}
and we recover the conservative Lorenz equations when $w=1$.
Finally, the conserved quantities for the system are given by the Casimir functions for (\ref{CLT}), namely 
\begin{equation}
\begin{array}{l}
\mathcal{C}=-2x^{2}+y^{2}+z(z+2 w),\\
\mathcal{W}=w.
\end{array}
\end{equation}

With this prerrequisites we can proceed with the formalism presented in Section 4. A cocommutator map compatible with the Lie algebra (\ref{lieL}) is found to be
\begin{equation}
\begin{array}{l}
\delta_\eta(X)= \eta\,(X\wedge Z+X \wedge W+\frac{1}{2}\,Y\wedge W), \\
\delta_\eta(Y)= \eta\,(X\wedge W+Y\wedge Z+Y\wedge W),\\
\delta_\eta(Z)=\delta_\eta(W)=0.
\end{array} \label{cocommutator}
\end{equation}
The dual Lie algebra $d_\eta$ obtained from the dual cocommutator map $\delta_\eta^*$ is given by
\begin{eqnarray}
&& [\hat x,\hat y]=0, \qquad [\hat x,\hat z]= \eta\, \hat x, \qquad [\hat x,\hat w]= \eta\, \hat x + \eta\, \hat y,  \nonumber \\
&& [\hat y,\hat z]= \eta\, \hat y, \qquad [\hat y,\hat w]= \frac{\eta}{2}\, \hat x + \eta\, \hat y, \qquad [\hat z,\hat w]= 0. \label{Pd}
\end{eqnarray}
This Lie algebra $d_\eta$ can be shown to be isomorphic to the direct sum of two copies of the two--dimensional real Lie algebra $[e_0,e_1]=e_1$. A four dimensional faithful representation $\rho$ of $d_\eta$ is given by the matrices
\begin{eqnarray}
\rho(\hat x)&=&\eta\,(e_1^3+e_1^4+e_2^4), \cr
\rho(\hat y)&=&\eta\,\left(\dfrac{1}{2}e_1^4+e_2^3+e_2^4\right), \cr
\rho(\hat z)&=&-\eta\,(e_1^1+e_2^2), \\
\rho(\hat w)&=&-\eta\,\left(e_1^1+\dfrac{1}{2}e_1^2+e_2^1+e_2^2\right). 
\nonumber
\end{eqnarray}
By using $\rho$, we construct the matrix group element
\begin{eqnarray}
&& \!\!\!\!\!\! \!\!\!\!\!\! \!\!\!\!\!\! \!\!\!\!\!\! D=\exp(x \rho(\hat x))\exp(y  \rho(\hat y)) \exp(z  \rho(\hat z)) \exp(w  \rho(\hat w))= \nonumber \\
&& \!\!\!\!\!\! \!\!\!\!\!\! =\left( 
\begin{array}{ccccccc}
\frac{1}{2} e^{-\left(w(1+\frac{1}{\sqrt{2}})+z\right) \eta } \left(1+e^{\sqrt{2} w \eta }\right) & 
-\tfrac{1}{2 \sqrt{2}}\,{e^{-\left(w(1+\frac{1}{\sqrt{2}})+z\right) \eta } \left(-1+e^{\sqrt{2} w \eta }\right)} & 
\eta\, x  & \eta \left(x+\tfrac{1}{2}y\right) \,\\
-\tfrac{1}{\sqrt{2}} \,{e^{-\left(w(1+\frac{1}{\sqrt{2}})+z\right) \eta } \left(-1+e^{\sqrt{2} w \eta }\right)}& 
\tfrac{1}{2} e^{-\left(w(1+\frac{1}{\sqrt{2}})+z\right) \eta } \left(1+e^{\sqrt{2} w \eta }\right) & \eta\, y  & \eta\, (x+y) \\
0 & 0 & 1  & 0 \\
0 & 0 & 0  & 1 \\
\end{array}
\!\! \!\! \!\! \right). \label{matrix}
\end{eqnarray}
As a consequence, the coproduct map arising from the group multiplication of two $D$ matrices is found to be
\begin{equation}
\begin{array}{l}
\Delta_\eta(x)=x_{1}+\dfrac{1}{2}e^{-\eta\, (z_{1}+w_{1})}\left[2 x_{2}\cosh \left(\tfrac{\eta}
{\sqrt{2}}\, w_{1}\right)-\sqrt{2}y_{2}\sinh \left(\tfrac{\eta}{\sqrt{2}}\, w_{1}\right)\right],\\
\Delta_\eta(y)=y_{1}+e^{-\eta\, (z_{1}+w_{1})}\left[y_{2}\cosh \left(\tfrac{\eta}
{\sqrt{2}}\, w_{1}\right)-\sqrt{2}x_{2}\sinh \left(\tfrac{\eta}{\sqrt{2}}\, w_{1}\right)\right], \\
\Delta_\eta(z)=z_1+z_2,\\
\Delta_\eta(w)=w_1+w_2.
\end{array}
\label{colo}
\end{equation}

Now, by applying again the algorithmic procedure introduced in~\cite{dualJPA}, we get the
following unique solution for the Poisson--Lie structure on the group $D$ that has $g^\ast$ as its linearization and is compatible with~\eqref{colo}: 
\begin{equation}
\{x,y\}_\eta=\dfrac{-1+e^{-2\eta(w+z)}-\eta^{2}(y^{2}-2\,x^{2})}{4 \eta }, 
\qquad \{x,z\}_\eta=\dfrac{y}{2}, \qquad \{y,z\}_\eta=x, \qquad \{w,\cdot\}_\eta=0. \label{DT}
\end{equation}
This is indeed an $\eta$--deformation of the $g^\ast\simeq gl(2)$ Lie--Poisson algebra~\eqref{CLT}, which is recovered in the $\eta\to 0$ limit, and the (deformed) Casimir functions for the Poisson algebra (\ref{DT}) are found to be 
\begin{equation}
\mathcal{W}_{\eta}=w,
\qquad
\mathcal{C}_{\eta}=
\frac{e^{-z \eta } \left(1+e^{2 \eta(w+z)  } \left[1+\left(-2 x^2+y^2\right) \eta ^2\right]-2 e^{z \eta } [1+w \eta  (1+w \eta )]\right)}{\eta ^2}.
\label{casL}
\end{equation}
Note that, as expected,
\begin{equation}
\lim\limits_{\eta\rightarrow 0}\mathcal{C}_{\eta}=-2x^{2}+y^{2}+z^{2}+2 z w=\mathcal{C}.
\end{equation}

Therefore, an integrable deformation of the integrable Lorenz system~\eqref{CL2} is obtained by considering the same Hamiltonian~\eqref{H1} and the Poisson--Lie brackets~\eqref{DT}, and reads
\begin{equation}
\begin{array}{ll}
\dot{x}=& y,\\
\dot{y}=&\frac{x}{2}\left(
4+\frac{1}{\eta}(e^{-2 \eta(z+w)}-1)+\eta\,(2x^{2}-y^{2})
\right)\\
\phantom{\dot{y}}=& x\, (2-w-z) +  x \left(x^2 - \tfrac{1}{2} \,y^2+ (w + z)^2\right) \eta -\frac{2}{3} x\, (w+z)^3\eta^2 + o[\eta^3] , \\[2pt]
\dot{z}=& x \, y ,
\end{array}
\label{CL2def}
\end{equation}
where the only deformed equation occurs for the $y$ variable, and the conserved quantities are given by the Hamiltonian and the Casimir functions~\eqref{casL}. Several periodic trajectories for this system are plotted in Figure 2, and are compared with the underformed Lorenz ones.

 
 

\begin{figure}[ht]
\setlength{\unitlength}{1mm}
\begin{picture}(140,66)(0,0)
\label{figure2}
\footnotesize{
\put(11,3){\includegraphics[scale=0.25]{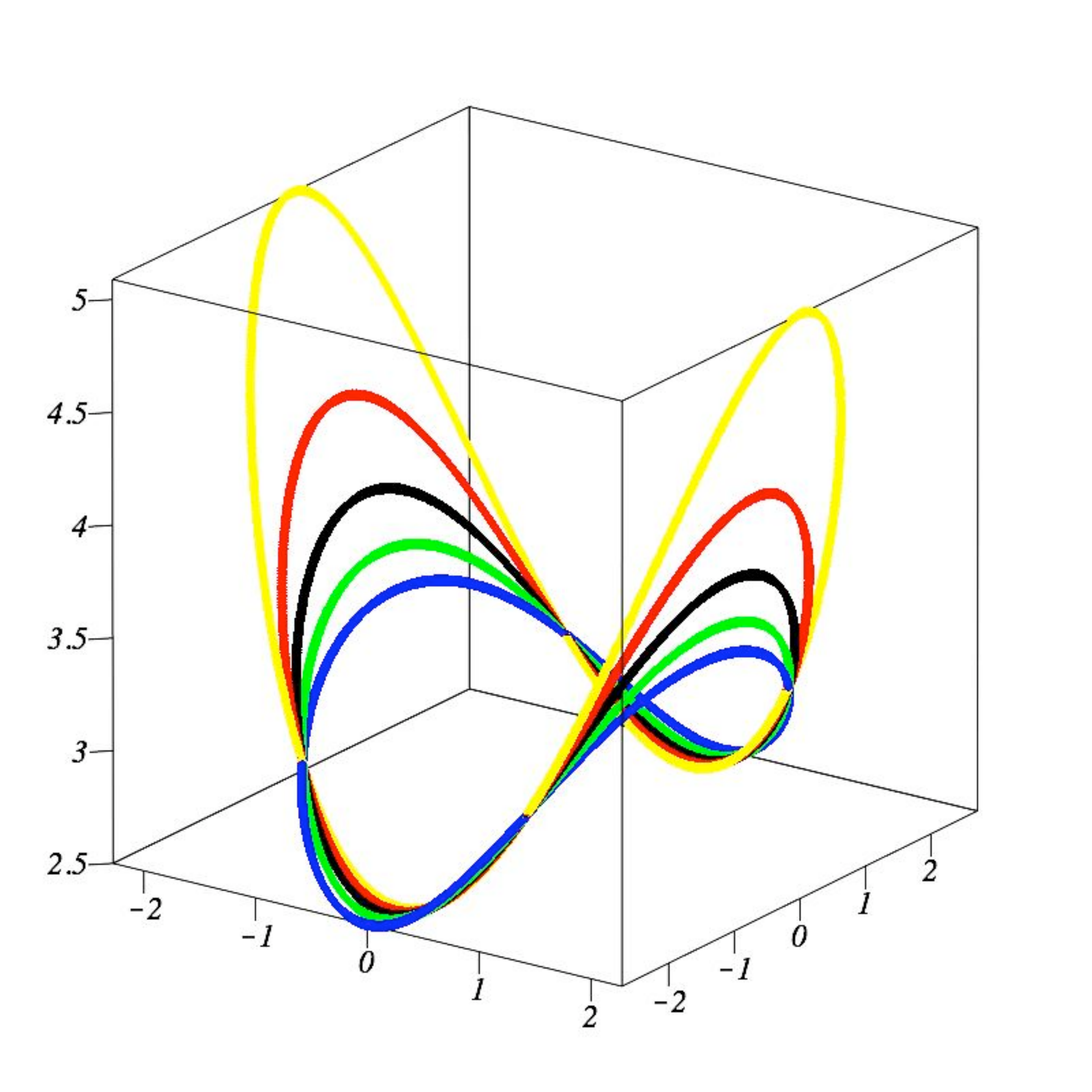}}
\put(38,6){\footnotesize $x$}
\put(68,12){\footnotesize $y$}
\put(13,36){\footnotesize $z$}
\put(11,1){(a)}
\put(85,3){\includegraphics[scale=0.25]{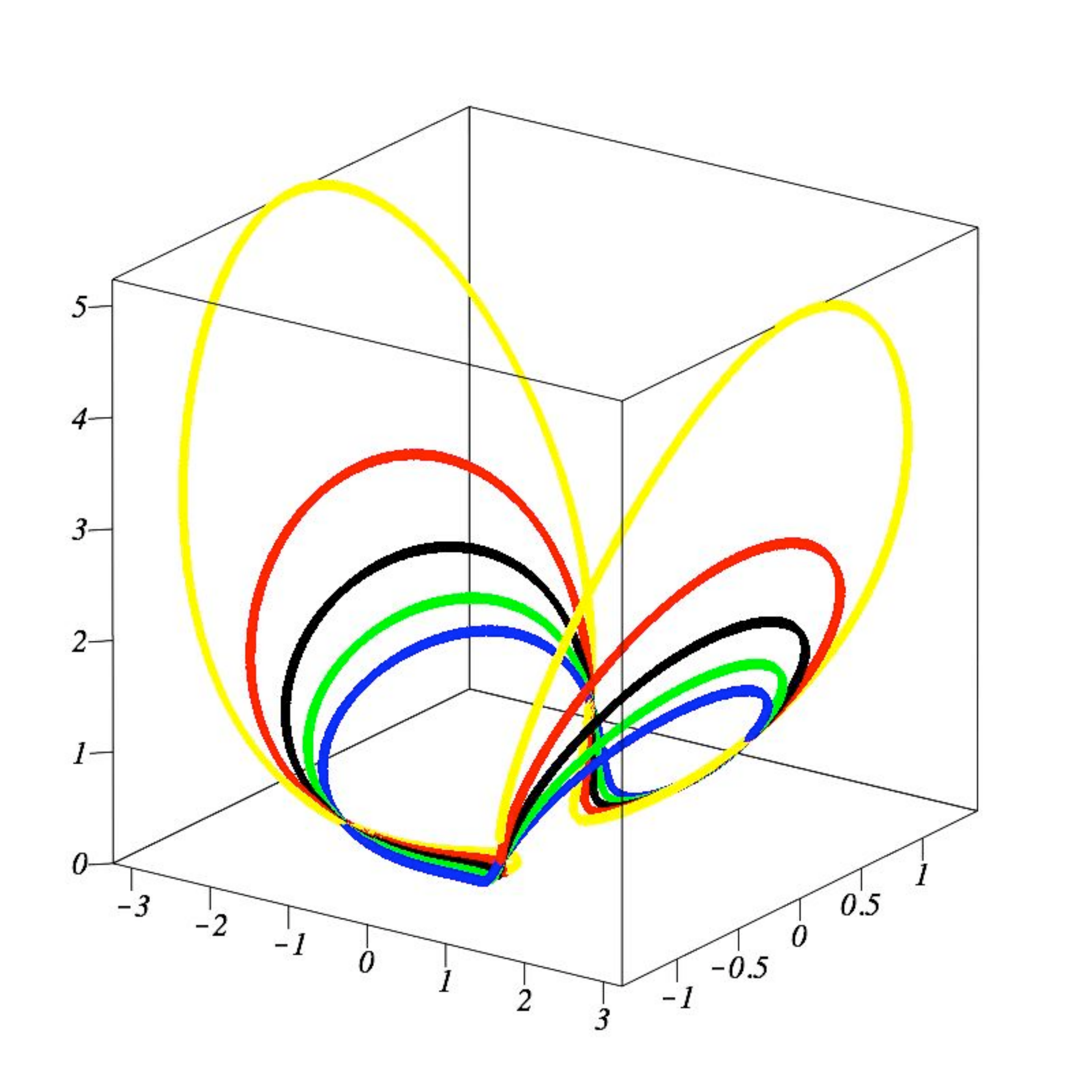}}
\put(112,6){\footnotesize $x$}
\put(142,12){\footnotesize $y$}
\put(87,36){\footnotesize $z$}
\put(85,1){(b)}
}
\end{picture}
\caption{ \footnotesize  
 (a) The closed trajectories of the conservative Lorenz system (\ref{CL2}) for the initial data $x(0)=1,y(0)=2,z(0)=3$ (black line) and of the deformed    
 conservative Lorenz system (\ref{CL2def}) with the same initial data, $w=1$ and $\eta=-0.1$ (blue line), $\eta=-0.05$ (green line),
 $\eta=0.05$ (red line), $\eta=0.1$ (yellow line). (b) The same figure as (a) but with initial data $x(0)=1,y(0)=-1,z(0)=0.5$. }
\end{figure}
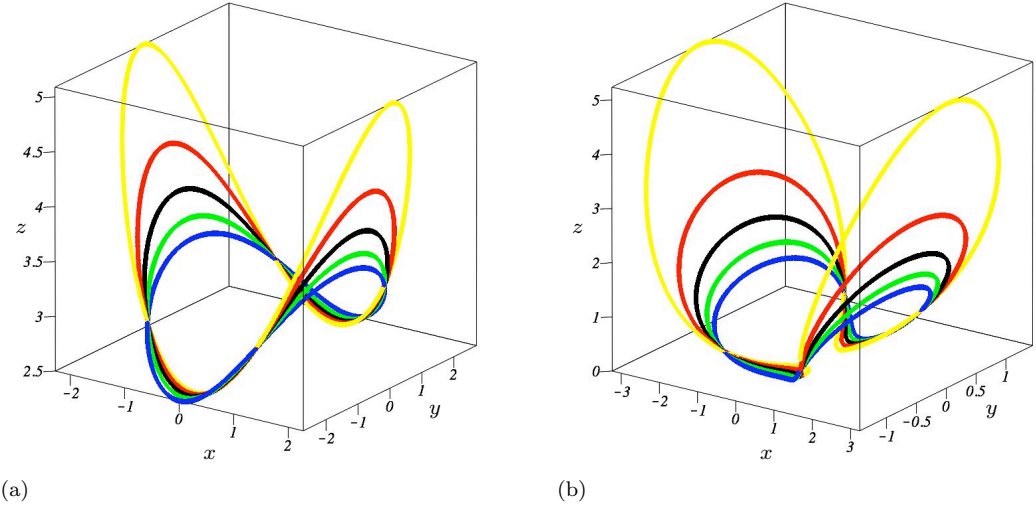


\subsection{Integrable coupled Lorenz system}


The undeformed Lie--Poisson coalgebra $({\cal{F}}(g^*),\{,\}_0,\Delta)$ is again defined through the primitive coproduct map~\eqref{coproducts}. Therefore, the $N=2$ coupled Lorenz Hamiltonian will be given by
\begin{equation}
\Delta(\mathcal{H})=2(\Delta(z)-\Delta(w))- [\Delta(x)]^{2},
\end{equation}
where the coproduct $\Delta$ is the undeformed one~\eqref{coproducts}. The corresponding equations of motion read
\begin{equation}
\begin{array}{l}
\dot{x}_{1}=y_1 , \\
\dot{y}_{1}= 2 x_1-(x_1+x_2)(w_1+z_1),\\
\dot{z}_{1}=y_1(x_1+x_2), \\[4pt]
\dot{x}_{2}= y_2,\\
\dot{y}_{2}= 2x_2-(x_1+x_2)(w_2+z_2),\\
\dot{z}_{2}= y_2(x_1+x_2),
\end{array}
\label{couplo}
\end{equation}
together with $\dot{w}_{1}=\dot{w}_{2}=0$ (see~\cite{Liu, He, Horikawa} for some instances of nonintegrable coupled Lorenz oscillators). Again, three additional  independent integrals of the motion in involution for this system are provided by the coalgebra symmetry: namely, the Casimir functions on each of the two copies of the Lie--Poisson algebra
\be
\mathcal{C}_1= -2x_1^2+y_1^2+z_1(z_1+2 w_1),
\qquad
\mathcal{C}_2= -2x_2^2+y_2^2+z_2(z_2+2 w_2),
\ee
together with the coproduct of the Casimir $\mathcal{C}$, which reads
\begin{equation}
\Delta(\mathcal{C})=-2 ({x_1}+{x_2})^2+({y_1}+{y_2})^2+({z_1}+{z_2}) (2 ({w_1}+{w_2})+{z_1}+{z_2}).\end{equation}

The $N$--dimensional integrable generalization of this system arises from the $N$-th coproduct map $\Delta^{(N)}$ of the Hamiltonian, namely
\begin{equation}
H^{(N)}:=\Delta^{(N)}(\mathcal{H})=2\left(
\sum\limits_{i=1}^{N}z_{i}-\sum\limits_{i=1}^{N}w_{i}
\right)-\left(\sum\limits_{i=1}^{N}x_{i}\right)^{2}.
\end{equation}
In this way the following $N$-coupled Lorenz system is obtained as
\begin{equation}
\begin{array}{l}
\dot{x}_{i}=y_{i} , \\
\dot{y}_{i}=2x_{i}-(z_i+w_i)\left(\sum\limits_{j=1}^{N}x_{j}\right), \qquad\qquad\qquad i=1,\dots,N\\
\dot{z}_{i}=y_{i}\sum\limits_{j=1}^{N}x_{j} ,
\end{array}
\label{coroNbis}
\end{equation}
and the ($2N -1 $) integrals of the motion (in involution between themselves and with respect to the Hamiltonian $H^{(N)}$) are given by~\cite{BR} the `one--site' integrals
\be
\mathcal{C}_i=-2x_i^2+y_i^2+z_i(z_i+2 w_i) , \qquad i=1,\dots,N
\qquad
\ee
together with the $k$--th order coproducts of the Casimir functions
\begin{equation}
C^{(k)}:=\Delta^{(k)}(\mathcal{C})=-2\left(\sum\limits_{j=1}^{k}x_{j}\right)^{2}+\left(\sum\limits_{j=1}^{k}y_{j}\right)^{2}
+\left(\sum\limits_{j=1}^{k}z_{j}\right)\left[
\sum\limits_{j=1}^{k}z_{j}+2 \sum\limits_{j=1}^{k}w_{j}
\right],
\quad
k=2,\dots, N.
\end{equation}
Note that the coupling between the 3D dynamical systems is induced by the non--linear $x^2$ term in the Hamiltonian.



\subsection{An integrable deformation}

By following the very same approach as for the R\"ossler system, the deformed coproduct map $\Delta_\eta$~\eqref{colo} acting on the Hamiltonian function  (\ref{H1}) gives rise to 
\be
\Delta_\eta(\mathcal{H})=2(\Delta_\eta(z)-\Delta_\eta(w))- [\Delta_\eta(x)]^{2},
\ee 
and by taking into account the explicit expressions for $\Delta_\eta$ we get
\be
\Delta(\mathcal{H})=2(z_{1}+z_{2}-w_{1}-w_{2})-\left[
x_{1}+\dfrac{1}{2}e^{-\eta\, (z_{1}+w_{1})}\left(2\, x_{2}\cosh \left(\tfrac{\eta}{\sqrt{2}}w_1\right)-\sqrt{2}\,y_{2}\sinh \left(\tfrac{\eta}{\sqrt{2}}w_1\right)\right)
\right]^{2} .
\ee

This Hamiltonian gives rise to the following (quite involved) deformation of the $N=2$ coupled Lorenz system
\begin{eqnarray}
\dot{x}_{1}&=& y_{1}+y_{1}\dfrac{\eta}{4}e^{-2\eta (w_{1}+z_{1})}\left[
(2x_{2}^{2}-y_{2}^{2})+2x_{1}e^{\eta (w_{1}+z_{1})}\left(
2x_{2}\cosh\left(\tfrac{\eta}{\sqrt{2}}w_1\right)-\sqrt{2}y_{2}\sinh \left(\tfrac{\eta}{\sqrt{2}}w_1\right)
\right)
\right]\notag\\
&& +y_{1}\dfrac{\eta}{4}e^{-2\eta (w_{1}+z_{1})}\left[
(2x_{2}^{2}+y_{2}^{2})\cosh(\sqrt{2}\,\eta\, w_{1})-2\sqrt{2}\,x_{2}\,y_{2}\sinh(\sqrt{2}\,\eta\, w_{1})
\right],\notag\\
\notag\\
\dot{y}_{1}&=&
\dfrac{x_{1}}{2\eta}\left(
-1+\eta (4+\eta(2x_{1}^{2}-y_{1}^{2}))
\right)\notag\\
&& +\dfrac{x_{1}}{2\eta}e^{-2\eta (w_{1}+z_{1})}
\left(
1+\eta^{2}(2x_{2}^{2}+y_{2}^{2})\cosh(\sqrt{2}w_{1}\eta)
+\eta^{2}(2x_{2}^{2}-y_{2}^{2}-2\sqrt{2}x_{2}y_{2}\sinh(\sqrt{2}w_{1}\eta))
\right)\notag\\
&& +\dfrac{1}{4\eta}e^{-3\eta (w_{1}+z_{1})}\left[
\left(
2x_{2}\cosh\left(\tfrac{\eta}{\sqrt{2}}w_1\right)-\sqrt{2}y_{2}\sinh\left(\tfrac{\eta}{\sqrt{2}}w_1\right)
\right)\left(1+e^{2\eta (w_{1}+z_{1})}(-1+\eta^{2}(6x_{1}^{2}-y_{1}^{2})) \right)
\right],\notag\\
\notag\\
\dot{z}_{1}&=& 
x_{1}y_{1}+\dfrac{1}{2}y_{1}e^{-\eta(w_{1}+z_{1})}\left(
2x_{2}\cosh \left(\tfrac{\eta}{\sqrt{2}}w_1\right)-\sqrt{2}y_{2}\sinh\left(\tfrac{\eta}{\sqrt{2}}w_1\right)
\right),
\notag\\
\notag\\
\dot{x}_{2}&=& y_{2}+\dfrac{1}{8\eta}e^{-2\eta(w_{1}+w_{2}+z_{1}+z_{2})}\left(
1+e^{2\eta (w_{2}+z_{2})}(-1+\eta^{2}(2x_{2}^{2}-y_{2}^{2}))
\right)\times \label{defLor2}\\
&& \times \left(
y_{2}-y_{2}\cosh(\sqrt{2}w_{1}\eta)+2\sqrt{2}x_{1}e^{\eta(w_{1}+z_{1})}\sinh\left(\tfrac{\eta}{\sqrt{2}}w_1\right)+\sqrt{2}x_{2}\sinh(\sqrt{2}w_{1}\eta)
\right),\notag\\
\notag\\
\dot{y}_{2}&=& 2x_{2}+\dfrac{1}{4\eta}e^{-2\eta (w_{1}+w_{2}+z_{1}+z_{2})}\cosh\left(\tfrac{\eta}{\sqrt{2}}w_1\right) 
  \left(1+e^{2\eta(w_{2}+z_{2})}(-1+\eta^{2}(2x_{2}^{2}-y_{2}^{2}))\right)\times\notag\\
&& \times \left(
2x_{1}e^{\eta (w_{1}+z_{1})}+2x_{2}\cosh\left(\tfrac{\eta}{\sqrt{2}}w_1\right)-\sqrt{2}y_{2}\sinh\left(\tfrac{\eta}{\sqrt{2}}w_1\right)
\right),\notag\\
\notag\\
\dot{z}_{2}&=& e^{-2\eta(w_{1}+z_{1})}\left(
x_{1}y_{2}e^{\eta(w_{1}+z_{1})}\cosh\left(\tfrac{\eta}{\sqrt{2}}w_1\right)+x_{2}y_{2}\cosh(\sqrt{2}\eta w_{1})
\right)\notag\\
&& -\dfrac{\sqrt{2}}{4}e^{-2\eta(w_{1}+z_{1})}\left(
4x_{1}x_{2}e^{\eta(w_{1}+z_{1})}\sinh\left(\tfrac{\eta}{\sqrt{2}}w_1\right)+(2x_{2}^{2}+y_{2}^{2})\sinh\left(\sqrt{2}w_{1}\eta\right)
\right).
\notag
\end{eqnarray}
By taking a power series expansion we get a first order deformation in $\eta$ of the coupled Lorenz system~\eqref{couplo} given by
\begin{equation}
\begin{array}{l}
\dot{x}_{1}=y_{1} +  x_2y_1(x_1+x_2) \eta  +o[\eta^2],\\[2pt]
\dot{y}_{1}=2x_{1}-(x_{1}+x_{1})(w_{1}+z_{1}) \\[2pt]
\quad\quad+ \left(
(w_{1}^{2}+z_{1}^{2}+2 w_{1}z_{1})(x_{1}+2x_{2})-\dfrac{y_{1}^{2}}{2}(x_{1}+x_{2})+\dfrac{w_{1}y_{2}}{2}(w_{1}+z_{1})+x_{1}(x_{1}^{2}+3 x_{1}x_{2}+2x_{2}^{2})
\right)\eta + o[\eta^2],\\
\dot{z}_{1}=y_1(x_1+x_2)-\dfrac{y_{1}}{2}(w_{1}(2x_{2}+y_{2})+2z_{1}x_{2})\eta + o[\eta^2]
\end{array}
\end{equation}
\begin{equation}
\begin{array}{l}
\dot{x}_{2}={y_2}-\dfrac{1}{2} {w_1} ({x_1}+{x_2}) ({w_2}+{z_2}) \eta +o[\eta^2],\\[2pt]
\dot{y}_{2}=(2 {x_2}-({x_1}+{x_2}) ({w_2}+{z_2})) \\
\quad\quad+\dfrac{1}{2} \left(
(x_{1}+x_{2})(2x_2^2-y_2^2+2(w_2+z_2)^2)+(w_2+z_2)(2 (x_1+2x_2)(w_1+z_1)+w_1 y_2)\right)\eta + o[\eta^2],\\[2pt]
\dot{z}_{2}=y_2({x_1}+{x_2})-\dfrac{1}{2} \left(
2 y_2 (w_1+z_1)(x_1+2x_2)+w_1 y_2^2+2 w_1 x_2 (x_1+x_2)
\right)\eta  + o[\eta^2].
\end{array}
\end{equation}

Finally, the three integrals of the motion for~\eqref{defLor2} are given by the coproduct $\Delta_\eta$~\eqref{colo} of the Casimir $C^{(2)}_\eta$ given in~\eqref{casL}, namely
\be
\begin{array}{l}
\Delta_\eta(C^{(2)}_\eta)=\dfrac{1}{\eta ^2} \, e^{-\eta\Delta_{\eta}(z)  } (1+e^{2 \eta(\Delta_{\eta}(w)+\Delta_{\eta}(z))  } \left[1+\left(-2 [\Delta_{\eta}(x)]^2+[\Delta_{\eta}(y)]^2\right) \eta ^2\right]-
\\
\qquad \qquad\qquad- 2 e^{ \eta \Delta_{\eta}(z)} [1+\eta\Delta_{\eta}(w)   (1+\eta\Delta_{\eta}(w)  )]),
\end{array}
\ee
together with the `one--site'\ deformed Casimirs
\be
\begin{array}{l}
\mathcal{C}^{(1)}_\eta=  \dfrac{e^{-z_1 \eta } \left(1+e^{2 \eta(w_1+z_1)  } \left[1+\left(-2 x_1^2+y_1^2\right) \eta ^2\right]-2 e^{z_1 \eta } [1+w_1 \eta  (1+w_1 \eta )]\right)}{\eta ^2},
\\
\mathcal{C}^{(2)}_\eta=  \dfrac{e^{-z_2 \eta } \left(1+e^{2 \eta(w_2+z_2)  } \left[1+\left(-2 x_2^2+y_2^2\right) \eta ^2\right]-2 e^{z_2 \eta } [1+w_2 \eta  (1+w_2 \eta )]\right)}{\eta ^2}.
\end{array}
\ee
Of course, since 
\be
\lim\limits_{\eta\rightarrow 0}\{\xi_{i},\Delta_{\eta}(\mathcal{H})\}_{\eta}=\{\xi_{i},\Delta(\mathcal{H})\},
\qquad
\xi_{i}=\{x_i,y_i,z_i\},
\ee
all these expressions define an integrable $\eta$-deformation of the $N=2$ coupled Lorenz system, and the generalization for arbitrary $N$ can be straightforwardly obtained through the $N$-th coproduct map that arises from~\eqref{colo} and from the recurrence~\eqref{rec}.



\section{Concluding remarks}

In this paper we have established a novel (and somehow unexpected) link between the theory of PL groups and the integrability problems for (coupled) ODEs. In fact, since Poisson--Lie groups can be thought of as deformations of Lie--Poisson coalgebras, this property can be used in order to construct integrable deformations of (coupled) systems of ODEs endowed with a Lie--Poisson Hamiltonian structure. The applicability of this statement is based on the fact that the explicit construction of the PL groups that deform a given Lie--Poisson algebra can be systematically obtained through the exponentiation of the associated Lie bialgebra structures (therefore, by making use of the Poisson version of the well--known `quantum duality principle'\ for quantum groups~\cite{Dri, STS}).  

One of the main advantages of the approach here presented relies on its algorithmic nature, that makes it amenable to be computerized  in a nearly complete way (see~\cite{dualJPA}).
In order to illustrate the method, in this paper we have worked out two specific integrable deformations of (coupled and uncoupled) R\"ossler and Lorenz systems. It is important to emphasize that other integrable deformations for these systems do exist. In fact, there will be as many of them as the number of possible Lie bialgebra structures associated to the initial Lie--Poisson  algebra that provides the Hamiltonian structure for the undeformed dynamical system (namely, the A$_{4,3}$ algebra~\eqref{LAconm} for the R\"osler system and the $gl(2)$ algebra~\eqref{lieL}  for the Lorenz one). 

Indeed, the very same approach can be applied to many other Hamiltonian systems of ODEs. Moreover, there also exists the possibility of generalizing this method in order to consider integrable systems with more than one Hamiltonian, either by deforming bihamiltonian Poisson pencils or by using the loop coproduct approach introduced in~\cite{Loop1,Loop2}. Finally it is worth mentioning that the analytic solution of coupled integrable systems with coalgebra symmetry is facilitated by the use of the so called `cluster'\  variables (see~\cite{BRcluster}), and this fact could shed some light on the dynamical significance of the R\"ossler and Lorenz couplings here obtained. Work on all these lines is in progress and will be presented elsewhere.



\section*{Acknowledgements}

This work was partially supported by the Spanish Ministerio de Econom\1a y Competitividad     (MINECO) under grant MTM2013-43820-P, by Junta de Castilla y Le\'on  under grant BU278U14 and by INFN--Sezione Roma Tre. 


\end{document}